\documentclass[a4paper,twoside]{article}

\usepackage{epsfig}
\usepackage{subcaption}
\usepackage{calc}
\usepackage{amssymb}
\usepackage{amstext}
\usepackage{amsmath}
\usepackage{amsthm}
\usepackage{multicol}
\usepackage{booktabs}
\usepackage{pslatex}
\usepackage{apalike}
\usepackage{float}
\usepackage{aliascnt}
\newaliascnt{eqfloat}{equation}
\newfloat{eqfloat}{h}{eqflts}
\floatname{eqfloat}{Equation}

\newcommand*{\ORGeqfloat}{}
\let\ORGeqfloat\eqfloat
\def\eqfloat{%
  \let\ORIGINALcaption\caption
  \def\caption{%
    \addtocounter{equation}{-1}%
    \ORIGINALcaption
  }%
  \ORGeqfloat
}

\usepackage[textsize=scriptsize, textwidth=7em]{todonotes}

\usepackage{changes}

\usepackage{SCITEPRESS}     

\begin{document}

\title{Phishing URL Detection Through Top-Level Domain Analysis: \\A Descriptive Approach}

\author{\authorname{Orestis Christou, Nikolaos Pitropakis, Pavlos Papadopoulos, Sean McKeown and William J. Buchanan}
\affiliation{School of Computing, Edinburgh Napier University, Edinburgh, United Kingdom}
\email{Christouorestis@gmail.com, \{N.Pitropakis, pavlos.papadopoulos, S.McKeown, B.Buchanan\}@napier.ac.uk}
}


\abstract{Phishing is considered to be one of the most prevalent cyber-attacks because of its immense flexibility and alarmingly high success rate. Even with adequate training and high situational awareness, it can still be hard for users to continually be aware of the URL of the website they are visiting. Traditional detection methods rely on blocklists and content analysis, both of which require time-consuming human verification. Thus, there have been attempts focusing on the predictive filtering of such URLs. This study aims to develop a machine-learning model to detect fraudulent URLs which can be used within the Splunk platform.
Inspired from similar approaches in the literature, we trained the SVM and Random Forests algorithms using malicious and benign datasets found in the literature and one dataset that we created. We evaluated the algorithms' performance with precision and recall, reaching up to 85\% precision and 87\% recall in the case of Random Forests while SVM achieved up to 90\% precision and 88\% recall using only descriptive features.}

\onecolumn \maketitle \normalsize \setcounter{footnote}{0} \vfill

\section{Introduction}
\label{sec:introduction}

\noindent The past few years have seen an outburst of high-impact breaches and issues, showing that sole reliance on traditional mitigation and prevention approaches is not ideal for providing adequate protection against such fluctuant environments. The Domain Name System (DNS), being one of the principal elements of the Web, is not only a prime target for attacks involving system downtime but is also used as a means for the execution of further and more complex social engineering and botnet attacks. In the report by Fouchereau and Rychkov (2019), 82\% of the companies undertaking their survey reported suffering from at least one DNS-related attack. The average number of attacks experienced per company was 9.45, placing the average cost of damages at \$1,000,000. From the \textit{dangerously diverse} and ever-growing threat landscape, the most ubiquitous DNS-related threat was phishing, with Malware, DDoS and Tunnelling following closely behind \cite{Fouchereau2019}. FireEye's recent report \cite{hirani2019global} on a grand-scale DNS hijacking attack by alleged Iran-based actors, for record manipulation purposes, reinforces this notion.

Adversaries do not have to be networking experts, nor possess knowledge of the underlying operation of the DNS to misuse it. To execute a successful phishing attack, the only thing an adversary needs to do is to select the right domain name to host their malicious website. Instead of merely choosing a generic and innocent-appearing name, the process of selecting the domain name may include techniques such as homograph spoofing or squatting \cite{kintis2017hiding,moubayed2018dns,nikiforakis2014soundsquatting}.

One of the most popular approaches for dealing with such websites is using blocklists~\cite{opendns2016phishtank}. It is simple and accurate as each entry in the blocklist is usually manually verified as malicious. The problem with the latter approach is that it requires frequent updating of the blocklist through constant scanning for new entries. Moreover, the systems creating these blocklists tend to have high operational costs, which lead to the companies requiring payment to access them. Usually, adversaries that utilise these malicious URLs do not keep them active for very long as they risk being detected and blocked. 

Machine Learning techniques use features extracted from the URLs and their DNS data to analyse and detect whether they are malicious or benign. Usually, methods which rely on the analysis of the content of such URLs come at a high computational cost. Blum et al. (2010) compute MD5 hashes of the main index of their webpages and compare them with the hashes of known phishing sites. In their work, they mention that this technique is easily bypassed just by obfuscating the malicious contents. This limitation constrains approaches to exclusively analysing the URL strings to classify the URLs \cite{blum2010lexical}. More recently, L{\'o}pez (2019), attempted to detect phishing by using \textit{Splunk} and taking into consideration only the use of typosquatting \cite{lopez2019metodos,nikiforakis2014soundsquatting} and homograph squatting. 

To the best of our knowledge, our work is the first attempt that takes into consideration all the forms of domain squatting and produces an automated mechanism to detect malicious phishing URLs, thus increasing the situational awareness of the user against them. The contributions of our work can be summarised as follows:
\begin{itemize}
\item The Machine Learning system is trained using descriptive features extracted from the URL strings without utilising host-based or bag-of-words features. The domain names come from real-world, known phishing domains (blocklist) and benign domain names (allowlist).
\item The popular classification algorithms SVM and Random Forests are used and compared empirically based on their performance.
\item The process is heavily automated as it relies on the Splunk software, thus it can easily be trained against new datasets to generate alerts when new malicious entries are detected.
\end{itemize}

The rest of the paper is organised as follows: Section 2 briefly describes the related literature with regards to phishing; Section 3 introduces our methodology while, Section 4 describes the results of our experimentation along with their evaluation. Finally, Section 5 draws the conclusions, giving some pointers for future work.

\section{Background and related work}

\subsection{Phishing}

\textit{Phishing} as a term did not exist until 1996 when it was first mentioned by \textit{2600} a popular hacker newsletter after an attack on \textit{AOL} \cite{ollmann2004phishing}. Since then, there has been an exponential increase in phishing attacks, with it becoming one of the most prevalent methods of cybercrime. According to Verizon (2019), phishing played a part in 78\% of all Cyber-Espionage incidents and 87\% of all installations of C2 malware in the first quarter of 2019 \cite{Verizon2019}. In the earlier report by Widup et al.~(2018), it is reported that \textit{78\% of people didn't click a single phish all year}, meaning that 22\% of people did click one. 

Moreover, only 17\% of these phishing campaigns were reported by users. It is also emphasised that even though training can reduce the number of incidents, \textit{phish happens} \cite{Verizon2018}. Since only a single e-mail is needed to compromise an entire organisation, protection against it should be taken seriously.

Cyber-criminals use phishing attacks to either harvest information or steal money from their victims through deceiving them with a reflection of what would seem like a regular e-mail or website. By redirecting the victim to their disguised website, they can see everything the victim inserts in any forms, login pages or payment sites. Cyber-criminals either copy the techniques used by digital marketing experts or take advantage of the fuss created by viral events to guarantee a high click rate. Vergelis and Shcherbakova (2019), reported a spike in phishing redirects to Apple's sites before each new product announcement \cite{Vergelis2019}.

\textbf{Regular phishing} attacks are usually deployed widely and are very generic, such that they can be deployed to target as many people as possible. A \textbf{Spear-Phishing attack}, instead, targets a specific individual, but requires that information be gathered about the victim prior to crafting a successful spear-phishing email.

A more advanced version of this attack is a \textbf{Whaling attack}, which specifically targets a company's senior executives to obtain higher-level access to the organisation's system. Targeted phishing attacks are increasingly gaining popularity because of their high success rates \cite{Krebs2018}.

\subsubsection{Attacks}

If the end-goal of a phishing attack is to ensure that the victim is ultimately redirected to the phishing website without being aware of it, then the adversary needs to use several techniques to guarantee that. One of those techniques is \textbf{URL hiding} , where the attacker obfuscates a malicious URL in a way that does not raise any suspicions and ultimately gets clicked on by the victim. One way to execute this would be to replace a valid URL link with a malicious one. \textbf{Shortened links} from services such as \textit{Bitly} can be used to obfuscate malicious links easily. There is no way to know the actual destination of an obfuscated link without visiting it.

\textbf{Homograph spoofing} is a method which depends on the replacement of characters in a domain name with other visually similar characters. An example of that would be to replace \textit{0} with \textit{o}, or \textit{I} with \textit{1} \cite{Rouse2019}. So, for a URL \textit{bingo.com} the spoofed URL would be \textit{b1ng0.com}. Characters from other alphabets such as Greek have also been used in the past for such attacks. The Greek \textit{o} character is visually indistinguishable from the English \textit{o} even though their ASCII codes are different and would redirect to different websites.

Content polymorphism is addressed, as well, using visual similarity analysis of the contents \cite{lam2009counteracting}.

\textbf{Typosquatting} targets common typographic errors in domain names. For example, an attacker could use the domain \textit{gooogle.com} to target users who incorrectly type \textit{google.com} or to trick them into clicking on a regular link. Moubayed et al. (2018), combat this issue using the K-Means Clustering Algorithm to observe the lexical differences between benign and malicious domains to extract features, and propose a majority voting system that takes into consideration the outputs of five different classification algorithms \cite{moubayed2018dns}. 

\textbf{Soundsquatting} leverages on the use of words that sound alike (homophones). Nikiforakis et al. (2014), show that for a domain \textit{www.test.com}, an adversary may use dot-omission typos (\textit{wwwtest.com}), missing-character typos (\textit{www.tst.com}), character-permutation typos (\textit{www.tset.com}), character-replacement typos (\textit{www.rest.com}) and character-insertion typos (\textit{www.testt.com}). They illustrate how they used Alexa's 
top one million domain list to create and register their soundsquatting domains, measuring the traffic from users accidentally visiting them. Through their research, they have proven the significance of taking into account homophone confusion through abuse of text-to-speech software when tackling the issue of squatting \cite{nikiforakis2014soundsquatting}.

\label{combosquatting}
\textbf{Combosquatting} is different from other approaches as it depends on altering the target domain by adding familiar terms inside the URLs. An example of this technique would be \textit{bankofscotland-live.com} or \textit{facebook-support.com}. Research performed by Kintis et al. (2017), shows a steady increase in the use of combosquatting domains for phishing as well as other malicious activities over time. It is also reported that combosquatting domains are more resilient to detection than typosquatting and that the majority of the combosquatting domains they were monitoring remained active for extended periods, sometimes exceeding three years, thus suggesting that the measures set in place to counter these are inadequate \cite{kintis2017hiding}.

\subsubsection{Suggested Defences}

The term Passive DNS (pDNS) refers to the indirect collection and archiving of DNS data locally for further analysis. In the early days of pDNS URL analysis~\cite{spring2012impact}, where privacy was still not considered to be an issue, the pioneer system for malicious domain detection through pDNS was Notos \cite{antonakakis2010building} with its reputation-based classification of domains. Notos extracts a variety of features from the DNS queries and creates a score for each entry to represent the likelihood of it being malicious. 

A similar approach is taken for EXPOSURE \cite{bilge2011exposure}, which is a large-scale pDNS analysis system developed using a gathered dataset of 100 billion entries. Bilge et al. (2011), differ in their approach by operating with fewer data compared to Notos. Khali et al. (2016) focus on the global associations between domains and IPs instead of looking at their local features, thus addressing any privacy issues as they only extract information relevant to their research from the gathered dataset \cite{khalil2016discovering}.

Okayasu and Sasaki (2015) compare the performance of SVM and quantification theory in a similar setting. SVM was proved to be superior in their comparison \cite{okayasu2015proposal}. 

Since the lexical contents of malicious URLs play a significant role in their victim's susceptibility, squatting detection should play a vital role in the identification of malicious URLs. Kintis et al.~(2017) use a joint DNS dataset comprised of 6 years of collected DNS records from both passive and active datasets that amount to a total of over 450 billion records. They found that the majority of combosquatting domains involve the addition of just a single token to the original domain \cite{kintis2017hiding}.

A novel approach is taken by Blum et al.~(2010), where URLs are classified without the need for host-based features. They found that lexical classification of malicious URLs can rival other conventional methods in accuracy levels \cite{blum2010lexical}. 

Lin et al.~(2013) propose a similar Machine Learning (ML) approach, which can detect malicious URLs by focusing on the URL strings. They use two sets of features to train their online learning algorithm: \textit{i)} Lexical features, which are extracted by taking the name of the domain, path, and argument of each entry, then using a dictionary to remove less useful words from them, and \textit{ii)} Descriptive features are static characteristics derived from the URLs such as total length or number of symbols \cite{lin2013malicious}.

Da Luz and Marques (2014), build upon the work of Notos \cite{antonakakis2011detecting} and EXPOSURE \cite{bilge2011exposure} to detect botnet activity using both host-based and lexical features. They give a comparison of the performance of the K-Nearest Neighbours, Decision Trees and Random Forests algorithms, showing that Random Forests performed significantly better~\cite{da2014botnet}.

Lin et al.~(2013) make the distinction between descriptive and lexical features by separating the features derived directly from the domain names strings and the features derived using their bag-of-words model. They use a combination of the Passive-Aggressive algorithm, to classify their dataset, and the Confidence Weighted algorithm, to alter the characteristic's weight, thus achieving more efficiency than other content-based models \cite{lin2013malicious}.

Our approach differentiates from these as it only emphasises the descriptive characteristics of URLs in order to observe and attempt to improve the performance of a model without taking into consideration host-based or lexical (bag-of-words) features. Moreover, none of the approaches in the literature explored the creation and usage of a model in a widely used platform such as Splunk, to provide alerting capabilities to automate the detection of malicious URLs.

\section{Methodology}

As malicious parties continue to abuse DNS to achieve their goals, the means of stopping them should also be constantly developed. When observing the literature historically, it can be deduced that modern approaches are becoming more focused on the detection of specific problems. Therefore, following the same trend, our work is assisted by the Splunk platform to train and use a classifier to detect phishing domains through their extracted descriptive features.

\begin{figure}[h!]
\centering
  \includegraphics[width=0.7\columnwidth]{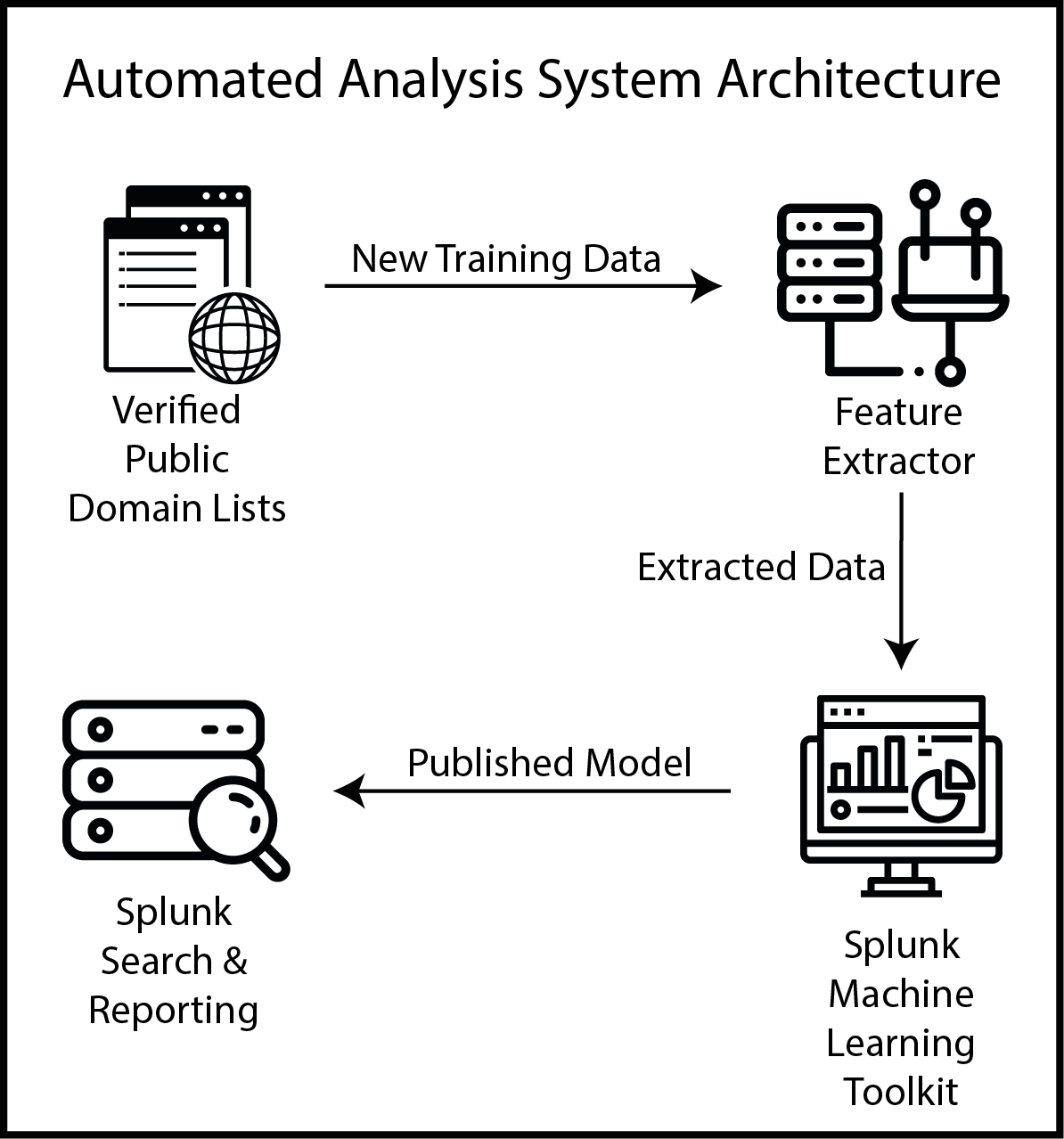}
  \caption{System Architecture Diagram}
  \label{fig:architecture}

\end{figure}

Figure~\ref{fig:architecture} illustrates the architecture of the proposed system. The technical specifications of our test environment are an Intel Core-i5 CPU with clock speed of 1.70GHz, and 6GiB RAM.

\subsection{Dataset Selection}

The quality of the prediction of a ML algorithm is strongly related to the quality of its training set. The Machine Learning approach requires a supervised learning algorithm, and therefore, the samples will need to be labelled as either \textit{benign} or \textit{malicious}. To reduce bias in our results, three tests were conducted using a total of six lists, three allowlists and three blocklists.

The first benign list was derived from TLDs in the Alexa's top one million domain list as of September 2019. This database contains one million entries of the most popular websites worldwide. As this list contains domains ranked by popularity, we verified their authenticity and content for the first five thousand domains from this list. We therefore populated our allowlist using the five thousand most popular Alexa domains. 

The first malicious list was created using Phishtank's active blocklist \cite{opendns2016phishtank}. The blocklist consists of more than four hundred thousand phishing domain entries and is continuously updated with active domains. Five thousand of those domains were selected to populate the blocklist. The two lists are joined, and overlapping domain names are removed to avoid creating any noise. More entries from the blocklists are included in the following tests. 

As the Alexa's database is not validated for potentially malicious entries like combosquatting domains, as proved in the related literature \cite{kintis2017hiding}, we turned our attention to established datasets used by the cyber security community. Therefore, the legitimate and malicious lists provided by Sahingoz et al. (2019) were used together in the second test. The legitimate list is reported to have been cross-validated and thus can be used at full scale \cite{sahingoz2019machine}. The phishing/legitimate URL set published in \textit{Phishtorm} \cite{marchal2014phishstorm} is used in the third test. The comparison of the performance of our feature selection using each of the different sets will be crucial in determining their relevance through the elimination of dataset bias from the algorithm's perspective.

\subsection{Analysis}

The previous sections have illustrated how miscreants can misuse DNS in their attempts to perform phishing attacks. From the knowledge extracted from the literature, a set of features was selected and extracted from the gathered pDNS data. The set of features allow for the classifier to divide the domain names into either benign or malicious.

\subsubsection{Feature Extraction}

The Splunk ML Toolkit has no functionality for extracting features from strings, and therefore, the features were extracted using Python libraries. The Pandas Python library was used to import the two datasets into Python for the extraction of the features.
 
We extracted a total of eighteen features from each domain name in the DNS dataset. Since the benign dataset did not include other details such as the TTL, certain types of features could not be derived. Therefore, more weight is given to analysing the lexical characteristics of our domains. The features are split into two groups: descriptive features and statistical features. The rationale behind the division of the features into these two categories is that descriptive features are variables derived directly from the domains while statistical features are derived from applying mathematical statistic operations on either the strings themselves or the descriptive features. 

\subsubsection{Descriptive features}

The descriptive features that were extracted from each domain name string are distinguished in the length, number of unique characters, numbers and symbols in the URL, in the domain, and in domain suffix. Malicious domain names tend to have a higher number of symbols or numbers than benign ones, either because of squatting or because they are randomly generated. Therefore, we extracted features to represent the number of symbols and numbers found in the different parts of the domain. 

Malicious domains also tend to be longer than benign ones \cite{moubayed2018dns}. However, after observing the entries in the phishing dataset, it was noticed that many entries would have longer subdomains but short domains. This would mean that even though they would look disproportionate, they would still be flagged benign as their length is similar to benign entries. To counter this issue, features were set to contain the length of each domain part.

The number of unique characters also differs in malicious URLs because legitimate website owners tend to choose simpler and easier to remember words for their URLs. Using this reasoning, the number of unique characters in the domain and suffix of each URL was used to populate the features. 
Moreover, since those unique characters are often numbers, features were selected to represent the number of numeric characters present in each URL. 

\subsubsection{Statistical Features}

Features to constitute the character continuity rate of the domain and suffix were selected. 
In general, as mentioned earlier, website owners tend to choose simpler names for memorisation purposes. Because simpler domain names are usually more expensive to buy, it is unlikely that the attackers will pay large sums for a domain that will most likely serve them for a short period. Lin et al.~(2013), use this idea to design the character continuity rate feature. To create this feature, the domain name is split into tokens of sequential characters based on their type (letter, number or symbol). Once the domain is split, the length of each token is measured and compared to the other tokens in its respective category. Then, the longest token for each character type is selected, and their total length is added together and divided by the total length of the token \cite{lin2013malicious}.

To detect randomised domain strings or at least detect randomisations within them, the features containing the Shanon entropy of the domain and suffix strings were selected \cite{lin2013malicious}. Shannon entropy H is calculated using the formula seen in Equation 4, where p\textsubscript{i} is the chance for a character i to appear in a given string \cite{marchal2012dnssm}.

\begin{eqfloat}
\centering
\begin{equation}
   H = \sum_{i} p_{i} log_{b} p_{i}
\end{equation}
\caption{Shannon Entropy}
\end{eqfloat}

In our scenario, p\textsubscript{i} is replaced with the count of different characters divided by the length of the string. 

Standard deviation and mean of the entropy of the two domain levels’ entropy were extracted to test if using a more median number would produce better results than using the initially extracted values of entropy.

\subsection{Training, Application and Alerting}

To train the algorithms, the training set containing all the features and the malicious and benign labels was exported. Splunk was set to monitor that file so that if any changes needed to be made to it, they could be updated in Splunk instantly. This experiment was split into Tests 1, 2 \& 3. Test 1 was performed using a set of ten thousand data entries from the Alexa and Phishtank datasets \cite{Alexadataset,opendns2016phishtank}. A 50/50 split was performed on the datasets for the training phase using Splunk to provide a better outlook of the algorithms' efficiency by using one half for training and the other for testing. 

The chosen algorithms for this experiment are SVM and Random Forests because of their reported performance. 
Support Vector Machine or SVM was developed based on Vapnik Chervonenkis' statistical learning theory in 1995 \cite{vapnik1995nature}. SVM is a linear model used to solve classification and regression problems. It can be applied to both linear and non-linear problems. Put simply; the algorithm separates data into classes using a line called the hyperplane.

Random Forests algorithm takes the same parameters as Decision Trees. It works by utilising multiple decision trees in order to improve the algorithm's robustness as well as its ability to generalise. The core idea of it is to use a multitude of trees that will output class predictions \cite{breiman2001random}.

Test 1.1 will examine the performance of the Random Forests algorithm with minimal changes to its default configuration of the parameters: infinite maximum depth, features and maximum leaf nodes, ten N estimators and two minimum samples per split. Test 1.2 will take a similar approach against the SVM algorithm, with slight alterations to the C and Gamma parameters which are set by default to 1 and 1/500 respectively.

Test 2 was separated into Test 2.1 and Test 2.2 to evaluate the SVM and Random Forests algorithms using a larger dataset of 70,000 data entries~\cite{sahingoz2019machine}. Likewise, Test 3 was divided into Test 3.1 and Test 3.2 to compare SVM and Random Forests with the Phishstorm dataset \cite{marchal2014phishstorm}. Test 4 serves as a \textit{what if} scenario that allows the comparison of all the available algorithms in the Splunk ML toolkit to see if there could have been alternatives not mentioned in the literature. 

After training the algorithm, Splunk will be configured to periodically fit the algorithm on a continuously monitored file so that any new entries are checked immediately for their maliciousness. If the algorithm is accurate enough, it will be configured for scheduled re-training to ensure that it is up to date with recently found phishing domain entries.

\section{Results \& Evaluation}

\subsection{Results}

The first algorithm to be tested was Random Forests in Test 1.1. In Table~\ref{tab:accuracytable}, the performance of the algorithm is evaluated. Moreover, the fine-tuning of the algorithm's parameters to achieve its full potential can also be seen. 

\begin{table*}[t]
\centering
\resizebox{\textwidth}{!}{%
\begin{tabular}{@{}lllllllllll@{}}
\toprule
\multicolumn{6}{|c|}{\textbf{Random Forests}} &  & \multicolumn{4}{|c|}{\textbf{SVM}} \\ \midrule
\textbf{Tests} & \textbf{N Estimators} & \textbf{Max Depth} & \textbf{Max Features} & \textbf{Precision} & \textbf{Recall} &  & \textbf{C} & \textbf{Gamma} & \textbf{Precision} & \textbf{Recall} \\
\textbf{1} & 10 & $ \infty$ & $ \infty$ & 0.87 & 0.86 &  & 1 & 1/18 & 0.89 & 0.87 \\
 & 10 & 10 & $ \infty$ & 0.89 & 0.86 &  & 1 & 1/50 & 0.83 & 0.83 \\
 & 10 & 10 & 2 & 0.89 & 0.87 &  & 10 & 1/18 & \textbf{0.90} & \textbf{0.88} \\
\textbf{2} & 10 & $ \infty$ & $ \infty$ & 0.84 & 0.84 &  & 1 & 1/18 & 0.76 & 0.76 \\
 & 1 & $ \infty$ & $ \infty$ & 0.81 & 0.81 &  & 10 & 1/18 & 0.77 & 0.77 \\
 & 10 & 10 & $ \infty$ & 0.80 & 0.80 &  & 100 & 1/18 & 0.78 & 0.77 \\
 & 10 & $ \infty$ & 2 & 0.84 & 0.84 &  & 100 & 1/500 & 0.79 & 0.77 \\
\textbf{3} & 10 & $ \infty$ & $ \infty$ & \textbf{0.85} & \textbf{0.85} &  & 1 & 1/18 & 0.79 & 0.79 \\
 & 1 & $ \infty$ & $ \infty$ & 0.83 & 0.83 &  & 100 & 1/18 & 0.81 & 0.81 \\
 & 10 & 10 & $ \infty$ & 0.83 & 0.83 &  & 100 & 1/100 & 0.81 & 0.80 \\
 & 10 & $ \infty$ & 2 & 0.85 & 0.85 &  & 100 & 1/500 & 0.80 & 0.79 \\ \bottomrule
\end{tabular}%
}
\caption{Accuracy of Tests}
\label{tab:accuracytable}
\end{table*}

Test 1.2 implemented the SVM algorithm against the same dataset as Test 1.1, with an initial precision \& recall of 0.89 and 0.87 respectively as shown in Table~\ref{tab:accuracytable}. The algorithm's performance increased immensely by simply using a larger C value to increase the hyperplane's flexibility. However, reducing the influence of the points placed far from the hyperplane resulted in a drop in accuracy. As shown in Table~\ref{tab:accuracytable}, a precision of 0.90 and a recall of 0.88 was achieved after the tweaking.

Test 2.1 evaluated the performance of the Random Forests algorithm using a larger dataset containing 70,000 entries. Once again, the features were tweaked until the perfect combination was found. As shown in Table~\ref{tab:accuracytable}, the performance of Random Forests peaked with a 0.84 precision and recall when 10 Estimators (Decision Trees) and an infinite max depth of nested statements were set. The alteration of the minimum samples per split had no impact on the overall performance, and therefore, it was kept to its default value. The decrease of the maximum number of features to consider per split negatively impacted the overall accuracy and thus was kept as the default value.

The same procedure as earlier is repeated using the SVM algorithm in Test 2.2. Using the default settings with a C of 1 and a Gamma of 1/18, the algorithm achieved a 0.76 precision and recall. In Table~\ref{tab:accuracytable}, it can be observed that this time, the peak performance was achieved using a C value of 100 and a Gamma of 1/500, with 0.79 precision and 0.77 recall achieved.

Table~\ref{tab:accuracytable} shows the performance of the Random Forests algorithm against the Phishstorm dataset \cite{marchal2014phishstorm} with 96,000 data entries in Test 3.1. Tweaking the algorithm parameters did not yield better results and thus the default configuration of the algorithm was kept.

Once more, the process is repeated using SVM in Test 3.2 (Table~\ref{tab:accuracytable}). When a C value of 100 was used the algorithm reached its peak precision \& recall rates of 0.81 respectively. Altering the Gamma value only reduced our rates and thus was kept to its default of 1/18.

After completing the experiments, the experimental setup of Test 3.1 was chosen to create a model in Splunk. The model allows for the fitting of the now trained algorithm into other datasets which are imported in Splunk. 

Splunk was configured to continuously monitor the input file to update any further additions or removals. An alert was created, which runs the query against the input file every hour and notifies the user if a new entry is flagged as malicious. The alert is then added to the triggered alerts. With this, the automated detection system is complete. Finally, the model is set to re-train itself using any new data added to the initial dataset.

\subsection{Evaluation}

The outcomes of Tests 1.1 \& 1.2 established that SVM performed slightly better than Random Forests. With Random Forests, the precision \& recall rates did not vary significantly when changing the algorithm's parameters. Decreasing the Gamma value in SVM however proved to have great significance on its results, indicating that even the furthest points from the hyperplane were of great importance. The evaluation of Tests 1.1 \& 1.2 should yield better results than the other tests due to the reduced number of entries. 

The full dataset containing 70,000 entries was used in Tests 2.1 \& 2.2 as the benign entries were already validated \cite{sahingoz2019machine}. Test 2.1 hit its peak performance without any tweaking, while Test 2.2 required a less straight hyperplane to do so. The evaluation results of SVM and Random Forests were not ideal for an automated filtration system but are still usable in our experiment. This time Random Forests was the predominant algorithm with a significant difference in precision \& recall. Overall both algorithms achieved lower rates, the sudden drop in accuracy of SVM indicates that Random Forests could be more ideal for a large-scale application. 

Using the Phishstorm dataset \cite{marchal2014phishstorm}, of 90,000 entries in Tests 3.1 \& 3.2 achieved very similar but slightly better results than Tests 2.1 \& 2.2. This means that the selected feature set is not dataset-biased and is robust when handling new data. Even though the achieved rates are not perfect, the model can still be used in the passive detection of malicious URLs. 

While SVM performed better than Random Forests in Test 1, Tests 2 \& 3 showed that Random Forests does not perform much differently when using a larger dataset, contrary to SVM's performance drop. Finally, Random Forests was selected for the final model simply because of the stability in the outcome it provides by utilising results from multiple decision trees.

After the model was created, it could be used as a standard search parameter in the original search and reporting app by Splunk and not just in the ML toolkit. This allowed for the easy creation of customised periodic checks for new malicious entries in the original dataset. In combination with a pDNS collector, this finalises the automated phishing URL detector. The feature for scheduled training could also be very easily implemented. 

\subsubsection{Feature Comparison}

After the \textit{publishing} of the model, the summary command in Splunk was used to evaluate the individual importance of the features used using the Random Forests algorithm.
Test 1 indicates that longer URLs (with more unique characters) tend to be malicious. These however may also be the differences between popular-expensive domains and unpopular-cheaper ones. The results derived from Test 2 were much more close, with the number of symbols in a domain name being the most important one. For the Test 3, the most important features are related with the total length, the number of the symbols and the character continuity rate in a domain name. An interesting observation can be made about the similarity in feature importance in Tests 2 \& 3 and their extreme contrast in performance with some features with Test 1 such as the number of symbols or character continuity rate. Another interesting observation is that the character continuity rate feature presented by Lin et al. (2013), had the third least importance of all the other features in Test 1, while in their experiment it ranked first. The count of numbers in the domain suffix was of no importance as there were no numbers in any of our URL suffixes \cite{lin2013malicious}.

\section{Discussion}

The precision and recall rates achieved here are good enough to indicate that the approach may be useful in real-life scenarios. They may not be ideal for an automated filtration system but can still provide a list of possibly malicious URLs, narrowing the list down and reducing the human input required to spot them. The sole use of descriptive features in a single classification may not be the correct approach if the goal is to achieve the efficiency levels required for an automated detection system to work. 

If the system had reached higher efficiency levels, then that would mean that the multi-lingual domain classification difficulties mentioned by Nikiforakis et al.~(2014), would be avoided. The purpose of selecting this specific approach was to further the trend of separate classification of features seen in the literature by not considering lexical features. It is an essential step towards understanding which groups of features work best together so that future developed multi-classifier systems are built knowing those relationships \cite{nikiforakis2014soundsquatting}.

The detection of phishing URLs is more challenging than the detection of botnet C\&C URLs as by their nature, phishing domain names attempt to mirror the appearance of benign domains. In retrospect, this model could have been a better fit for detecting randomly generated C\&C domains as their randomisation would lead to higher entropy values, longer URLs and would have used a multitude of unique characters and symbols. This would yield higher precision and recall rates which would, in turn, produce more accurate alerts once the model was \textit{published} in the Splunk search and reporting app. For the detection of phishing domains, more focus should have been given towards features explicitly targeting the detection of squatting domains \cite{moubayed2018dns}.

This difference in feature importance between the three tests as well as from other approaches in the literature suggests that the quality of the datasets used can entirely change which features will be more critical for the classification. It is concerning how a feature such as the character continuity Rate, had so little importance in the first experiment while in Tests 2 \& 3 and in the literature \cite{mamun2016detecting}, it was one of the most important. In the second approach taken by \cite{da2014botnet}, their Shannon entropy features are of similar importance to Shannon entropy of the domain and suffix strings in Test 1; high in the third level domain and low in the second level domain. However, in their first approach, which used a different dataset, they differed completely, with both features holding no importance. The positive evaluation of the Unique Characters features by \cite{moubayed2018dns}, mirrored their performance in the first test. The domains' length and symbol count features proved to be the most important in Tests 2 \& 3 but not the count of unique characters; homograph spoofed and typosquatted domains are the likely culprits responsible for this.

As demonstrated, it is possible to use the trained model to make predictions against new data within the Splunk Search and Reporting app. Though with the current model, it would be ill-advised to do so as the surge of false positive and false negative alerts generated would cause more harm than good. 

It would be best, however, if a classifier with higher accuracy was used.

\subsection{Attacks against Machine Learning}

The best method of evading detection from a ML algorithm is to just use an expensive domain name, meaning that a wealthy assailant can operate relatively unhindered. For this reason, a lexical analysis system should never be relied upon exclusively, but should be incorporated into a larger system.

The ML system is at its most vulnerable during training \cite{pitropakis2019taxonomy}, as that is where human error can thrive. If a malicious entry is included in the benign training dataset then not only will that particular entry not be detected later, but other malicious URLs with similar characteristics may also escape detection. A malicious entry in a allowlist (poisoning) has much more potential to cause damage than a benign entry in a blocklist.

A creative adversarial approach to invalidate a ML system and avoid detection would be to buy a swarm of malicious names with similar features and associate them with malicious activity. In time, after they are included in blocklists because their similarities are known to the adversary, it would be easier to select domain names which would bypass detection as they would have a hand in the training. Although in a smaller scale, this method would not have much accuracy, it is always a possibility in a grandiose cyber-warfare scenario. However, with so many resources, it would be much easier to simply buy an expensive domain.

\section{Conclusions and future work}

As adversaries keep inventing different means of abusing the DNS, the only certainty is that Machine Learning will continue to play a vital role in the future of malicious URL filtering. In this work, the descriptive features derived from benign and malicious domain name datasets were used to make predictions on their nature using the Random Forests and SVM algorithms. The final precision and recall rates produced when only using descriptive features and not considering host-based features were up to 85\% and 87\% for Random Forests and up to 90\% and 88\% for SVM respectively. Those results play a vital role in the understanding of the operational relationship between features and thus contribute knowledge into the correct grouping of features and the creation of multi-model classifiers. The features were found to have significantly different impact factors than some other cases in the literature, proving the importance of placing great care in the selection of training datasets. After the model was finalised and fine-tuned, it was \textit{published} in the Splunk Search and Reporting app where it was used against new data to generate alerts. This was the final step towards the automation of the detection process. Scheduled training was also configured using Splunk, furthering the system's autonomy. 

There are several research pathways which can be undertaken to improve the performance of this system, some being parallel to and some being stemming from the existing literature. A great addition to our methodology would be the use of a dataset composed of real-world passive DNS data for the training phase that would allow for the generation of more features, thus leading towards the elimination of noise. As we have a passive DNS infrastructure under development, we plan in the near future to make use of a higher volume of real-world data as training datasets, which would lead to the further improvement of our model.

\bibliographystyle{apalike}
{\small
\bibliography{example}}

\end{document}